\documentclass[]{aastex}
\usepackage{emulateapj5}

\usepackage{amssymb}
\usepackage{amsmath}
\usepackage{natbib}
\usepackage{epsfig}
\usepackage{epsf}
\usepackage{color}
\definecolor{red}{rgb}{0.7,0,0}
\definecolor{blue}{rgb}{0,0,0.7}

\def\X1859{XTE~J$1859$+$226$}
\def\ergcms{\ensuremath{\mathrm{erg}\,\mathrm{cm}^{-2}\,\mathrm{s}^{-1}}}
\def\integral{\textit{INTEGRAL}}

\def\chisq{\ensuremath{\chi^2_\nu}}
\def\cygx{Cyg~X-1}

\shorttitle{State dependence of the Cyg X-1 MeV polarization}
\shortauthors{Rodriguez et al. }
\begin{document}
\title{Spectral state dependence of the 0.4--2 MeV polarized emission in Cygnus~X-1 seen with INTEGRAL/IBIS, and links with the AMI radio data}

\author{
J\'er\^ome Rodriguez\altaffilmark{1},  Victoria Grinberg\altaffilmark{2,3}, Philippe Laurent\altaffilmark{4}, Marion Cadolle Bel\altaffilmark{5},  
 Katja Pottschmidt\altaffilmark{6,7}, Guy Pooley\altaffilmark{8},  Arash Bodaghee\altaffilmark{9},
 J\"orn Wilms\altaffilmark{2} \and Christian Gouiff\`es\altaffilmark{1}
 }

\altaffiltext{1}{Laboratoire AIM, UMR 7158, CEA/DSM - CNRS - Universit\'e Paris Diderot, IRFU/SAp, F-91191 Gif-sur-Yvette, France. \email: jrodriguez@cea.fr}
\altaffiltext{2}{Dr. Karl Remeis-Sternwarte and Erlangen Centre for Astroparticle Physics, Friedrich Alexander Universit\"at Erlangen-N\"urnberg, Sternwartstr. 7, 96049 Bamberg, Germany}
\altaffiltext{3}{Massachusetts Institute of Technology, Kavli Institute for Astrophysics, Cambridge, MA 02139, USA}
\altaffiltext{4}{Laboratoire APC, UMR 7164, CEA/DSM CNRS -- Universit\'e Paris Diderot, Paris, France}
\altaffiltext{5}{LMU, Excellence Cluster ``Universe'', Boltzmannstrasse 2, 85748 Garching, Germany}
\altaffiltext{6}{CRESST, University of Maryland Baltimore County, 1000 Hilltop Circle, Baltimore, MD 21250, USA}
\altaffiltext{7}{NASA Goddard Space Flight Center, Astrophysics Science Division, Code 661, Greenbelt, MD 20771, USA}
\altaffiltext{8}{Astrophysics, Cavendish Laboratory, J J Thomson Avenue, Cambridge CB3 0HE, UK}
\altaffiltext{9}{Georgia College \& State University, Dept.\ of Chemistry, Physics, and Astronomy, CBX 82, Milledgeville, GA 31061, USA}


\begin{abstract}
 Polarization of the $\gtrsim 400$~keV hard tail of the microquasar Cygnus X-1 has been independently reported by \integral/IBIS, and 
 \integral/SPI and interpreted as emission from a compact jet. These conclusions were, however,
based on the accumulation of all \integral\ data regardless of the spectral state. We utilize additional \integral\ exposure accumulated 
until December 2012, and include  the AMI/Ryle (15 GHz) radio data in our study. We separate the observations into hard, soft, and 
intermediate/transitional states and detect radio emission from a compact jet in hard and intermediate states, but not in the soft.  
The 10--400\,keV \integral\ (JEM-X and IBIS) state resolved spectra are well modeled with thermal Comptonization and reflection 
components. We detect a hard tail in the 0.4--2~MeV range for the hard state only. We extract the state dependent polarigrams of Cyg X-1, 
which all are compatible to no or undetectable level of polarization except in 400--2000\,keV range in the hard state where the polarization 
fraction is $75\pm32 \%$ and the polarization angle $40\fdg0\pm14\fdg3$. An upper limit on the 0.4--2\,MeV soft state polarization 
fraction is 70\%. Due to the short exposure, we obtain no meaningful constraint for the intermediate state. The likely detection 
of a $>400$\,keV polarized tail in the hard state, together with the simultaneous  presence of a radio jet, reinforce the notion of a compact 
jet origin of the $>400$\,keV emission.
\end{abstract}

\section{Introduction}
Black hole binaries (BHB) transit through different ``spectral states'' during their outbursts. The two canonical ones are the soft state (also known as ``high'' state, hereafter HSS) and the hard state (also known as ``low'' state, hereafter LHS). 

In the HSS, the emission
is dominated by a bright and warm ($\sim1$\,keV) accretion disk, the level of variability is low, and the power density spectrum
 is power-law like.  Little or no radio emission is detected in this state, which is interpreted as an evidence for an absence of  jets. 

 In the LHS, the disk is colder ($\leq0.5$\,keV) and might be truncated at a large distance from the accretor 
 \citep[see however][for examples of results contradicting this picture]{reynolds13}. The X-ray spectrum shows a strong power-law like component extending up to hundreds 
 of keV, usually with a roll-over at typically 50--200\,keV.  The level of rapid variability is higher than in the HSS and the power density spectrum s
shows a band limited noise component, sometimes with quasi-periodic oscillations (QPOs) with frequencies in the range $\sim0.1$--10\,Hz. A so-
called ``compact-jet'' is detected mainly through its emission in the radio to infrared domain \citep[e.g.,][]{corbel13} where the spectrum can be 
modeled by $F_\nu \propto \nu^{-\alpha}$ where $\alpha\leq0$ up to a break frequency $\nu_{\mathrm{break}}$ that usually lies in the infrared 
domain \citep{Corbel02a,rahoui11}. Above this break, the spectrum is a typical synchrotron spectrum
 with $\alpha\ge0.5$.

Intermediate states (IS) exist between the HSS and LH \citep[see, e.g.,][for reviews and more detailed state classifications]{remillard06,homan05b,fender09}.  Transitions between states indicate drastic changes in the properties of the flows and thus are crucial for the understanding of accretion-ejection connection. During a transition from the LHS to the HSS, the compact jet is quenched \citep[e.g.][]{fender99,mickael_1743}; discrete, sometimes superluminal ejections occur, the level of X-ray variability drops abruptly, and QPOs, if present in the given source, change types before disappearing \citep[see, e.g.,][and references therein for both a description of the different types of QPOs and a possible
 theoretical interpretation of their origin and link to the spectral states]{varniere11,varniere12}. 

The BHB Cygnus X-1 (Cyg X-1) was discovered in 1964 \citep{Bowyer65}.  It is the first Galactic source known to host a black hole \citep{Bolton75} as the primary, and recent estimates led to a black hole mass of $\mathrm{M}_{\mathrm{BH}}=14.8\pm1.0\,{\mathrm{M}}_\odot$ \citep{Orosz11}. The donor star is the O9.7 Iab star HDE 226868 \citep{Bolton72b,Walborn73} with $\mathrm{M}_{\mathrm{HDE}}=19.2\pm1.9\,M_\odot$ \citep{Orosz11}\footnote{See also \citet{ziolkowski14} who obtain somewhat different values.}. The systems orbital period is 5.6\,d \citep{Webster72, Brocksopp99, Gies03}, it has an inclination $i = 27\fdg1 \pm 0\fdg8$ on the plane of the sky \citep{Orosz11}, and is located at a distance of $d = 1.86 \pm 0.12$\,kpc from Earth \citep{Reid11, Xiang_2011}.

Cyg X-1 can be found in both the LHS and the HSS; up to 2010, it was predominantly in the LHS.  Since then, the behavior has changed and it spends most of its time in the HSS \citep{grinberg2013}.  It sometimes undergoes partial (``failed'') transitions from the LHS to the HSS, and can be found in a transitional or intermediate state \citep[e.g.,][]{Pottschmidt03b}. The detection of compact relativistic jets in the LHS \citep{Stirling01} places
Cyg X-1 in the family of microquasars. It is one of the few microquasars known to have a hard tail extending to (and beyond) the MeV range \citep[e.g.,][]{mcconnell00}. 

The presence of the MeV tail has recently been confirmed with the two main instruments onboard the ESA's INTErnational Gamma-Ray Astrophysics Laboratory (\integral): the Imager onBoard the \integral Satellite \citep[IBIS;][] {ubertini03} and the SPectrometer on \integral\ \citep[SPI;][]{vedrenne03}. \citet{marion04} and \citet[][; LRW11]{laurent11} detect it with IBIS, and \citet[][; JRC12]{jourdain12} with SPI.  Utilizing photon that Compton scattered in the upper plane of IBIS, \citep[ISGRI;][]{lebrun03}, and are absorbed in its lower plane PICsIT \citep[sensitive in the 200\,keV--10\,MeV energy band]{labanti03}, we have shown that the $\geq400$\,keV emission of Cyg X-1 was polarized at a level of about 70\%. We obtained only a 20\% upper limit on the degree of polarization at lower energies (LRW11). This result was independently confirmed by JRC12 using SPI data. Both studies obtained compatible results for the properties of the polarized emission, reinforcing the genuineness of this discovery. Both teams also suggested that the polarized emission was due to synchrotron emission coming from a compact jet. 

The presence of polarized emission and the energy-dependency of the polarization have profound implication on our understanding of accretion and ejection processes. It can help to distinguish between the different proposed emitting media (Comptonization corona vs.\ synchrotron-self Compton jets) in microquasars and provide important clues to the composition, energetics and magnetic field of the jet.  

A problem of both studies, however, is that they accumulated all \integral\ data available to them, regardless of the spectral state and radio (and thus jet) properties of the source. In this paper, we separate the whole \integral\ data set (accumulated up to December 2012) into different spectral states and study the properties of the state resolved broad band 10--2000\,keV emission of Cyg X-1. We also utilize the Ryle/AMI radio data to determine the state dependent level of 15\,GHz jet emission. In \S2, we describe the data reduction and in particular the so-called ``Compton mode'' that allowed us to discover and measure the polarization in Cyg X-1 (LRW11). We carefully separate the data into spectal state following a procedure based based on the classification of \citet{grinberg2013} that is described in \S\ref{sec:asmstate}. The results of both, the long term 15\,GHz radio monitoring and \integral\ state resolved analysis, are presented in \S4. We discuss the implications of our analysis in \S5 and summarize our results in \S6.

\section{Observations and data reduction}

\subsection{Standard data reduction of the JEM-X and IBIS/ISGRI data}

\cygx\ has been extensively observed by \integral\ since the launch of the satellite; preliminary results of the very first observations are described by \citet{katja03}.  In LRW11, we first considered all uninterrupted \integral\ pointings, also called ``science windows'' (ScWs), where \cygx\ had an offset angle of less than $10^\circ$ from the center of the field of view. We then removed all ScWs from the performance verification phase so that our analysis covered the period from 2003 March 24 (MJD 52722, satellite revolution [rev.] 54) to 2010 June 26 (MJD 55369, rev.\ 938). We further excluded ScWs with less than 1000\,s of good ISGRI and PICSiT time. This resulted in a total of 2098 ScWs. Polarization was detected when stacking this sample (LRW11).

In the present work, we added to the LRW11 sample all observations belonging to our Cyg X-1 \integral\ monitoring program (PI J. Wilms) made until 2012 December 28 (MJD 56289.8, rev.\ 1246). We applied the same 
filtering criteria to select good ScWs. We consider both, the data of the low energy detector of IBIS, ISGRI \citep{lebrun03} and of the X-ray monitors JEM-X \citep{lund03}. ISGRI is sensitive in the $\sim$18--1000\,keV energy range, although its response falls off rapidly above a few hundred\,keV. The two JEM-X units cover the soft X-ray (3--30\,keV) band.
All data were reduced with the \texttt{Off Line Scientific Analysis (OSA)} v.10.0 software suite and the associated updated calibration \citep{Caballero2012}, following the methods outlined by \citep{rodrigue08_1915b}. 

The brightest and most active sources of the field, Cyg X-1, Cyg X-2, Cyg X-3, 3A 1954+319, and EXO 2032+375, were subsequently taken into account in the extraction of spectra and light curves.  The Cyg X-1 spectra of each ScW were extracted with 67 spectral channels. All ScWs belonging to the same state were then averaged to produce one ISGRI spectrum for each spectral state. The spectral classification is described in \S\ref{sec:asmstate} \citep[see][for the details of the method]{grinberg2013}.  A systematic error of 1\% was applied for all spectral channels. Low significance channels were rebinned when necessary.

The JEM-X telescopes have a much smaller field of view than IBIS. The number of ScWs during which Cyg X-1 can be seen by these instruments is therefore smaller. Since rev.\ 983 (MJD 55501) both units are on for all observations. Before rev.\ 983, both units were used alternately, so that the number of ScWs and time coverage is not the same. We derived images and spectra in the standard manner following the cookbook, separating the data into states following the same criteria as for the IBIS data.  We applied systematic errors of 2\% to all spectral channels of both units.

\subsection{The Compton Mode}
\label{sec:comptonmode}
Thanks to its two position-sensitive detectors ISGRI and PICsIT, 
IBIS can be used as a Compton polarimeter \citep{lei97}. The concept behind a Compton polarimeter utilizes the polarization dependence of the differential cross section for Compton scattering, 
\begin{equation}
\frac{d\sigma}{d\Omega} = \frac{r_0^2}{2}\left(\frac{E'}{E_0}\right)^2\left(\frac{E'}{E_0}+\frac{E_0}{E'}-2 \sin^2\theta \cos^2\phi \right)
\end{equation}
where $r_0$ is the classical electron radius, $E_0$ the energy of the incident photon, $E'$ the energy of the scattered photon, 
$\theta$ the scattering angle, and $\phi$ the azimuthal angle relative to the polarization direction. Linearly polarized photons 
scatter preferentially perpendicularly to the incident polarization vector. Hence by examining the scattering angle distribution 
of the detected photons (also referred to as polarigrams)
\begin{equation}
N(\phi)=S[1+a_0\cos(2(\phi-\phi_0))]
\label{eq:azimuth}
\end{equation}
where $S$ is the mean count rate, one can derive the polarization angle $PA$ and polarization fraction $\Pi$. 
With $PA = \phi_0 - \pi /2 + n \pi$, equation~\ref{eq:azimuth}
becomes  $N(\phi)=S[1-a_0\cos(2(\phi-PA))] [2\pi]$. The polarization angle therefore corresponds to the minimum of $N(\phi)$. 
The polarization fraction is  $\Pi= a_0/a_{100}$, where $a_{100}$ 
is the amplitude expected for a 100\% polarized source, obtained from Monte-Carlo simulations of the instrument. 
 \citet{forot08} obtained $a_{100}=0.30 \pm 0.02$, hence a $\sim6.7\%$ systematics uncertainty that should be 
 taken into account in the derivation of $\Pi$. Recent simulations allowed us to reduce
 the uncertainty on $a_{100}$ to $\sim3\%$ which is small compared to the statistical errors on $N(\phi)$ (see below), and  
 is therefore neglected.
 
 To measure $N(\phi)$, we followed the procedure described by \citet{forot08} that led to the successful detection of the polarized signal from the Crab nebula. We consider events that interacted once in ISGRI and once in PICsIT. These events are automatically selected on board through a time coincidence algorithm. The maximum allowed time window was set to 3.8\,$\mu$s during our observations.  To derive the source flux as a function of $\phi$, the Compton photons were accumulated in 6 angular bins, each with a width of $30^\circ$ in azimuthal scattering angle. To improve the signal-to-noise ratio in each bin, we took advantage of the $\pi$-symmetry of the differential cross section (Eq.~\ref{eq:azimuth}), i.e., the first bin contains the photons with $0^\circ\leq\phi<30^\circ$ and $180^\circ\leq\phi<210^\circ$, etc. Chance coincidences, i.e., photons interacting in both detectors, but not related to a Compton event, were subtracted from each detector image following the procedure described by \citet{forot08}. The derived detector images were then deconvolved to obtain sky images. The flux of the source in each $\phi$-bin was then measured by fitting the instrumental PSF to the source peak in the sky image.

The uncertainty of $N(\phi$) is dominated by statistic fluctuations in our observations that are background dominated. Confidence intervals on $a_0$ 
and $\phi_0$ are not derived by a $N(\phi)$ fit to the data, but through a Bayesian approach following \citet{forot08},  based on the work  
presented in \citet{vaillancourt06}. The applicability of this method was recently thoroughly discussed in \citet{maier14}.
In this computation, we suppose that all real polarization angles and fractions have an uniform probability distribution  \citep[non-informative prior 
densities; ][]{quinn12, maier14} and that the real polarization angle and fraction are $\phi_0$ and $a_0$. We then need the probability density 
distribution of measuring $a$ and $\phi$ from $N_\mathrm{pt}$ independent data points in $N(\phi)$ over a period $\pi$, based on Gaussian 
distributions for the orthogonal Stokes components \citep{vaillancourt06, forot08,maier14}:
\begin{multline}
dP(a,\phi) = \frac{N_\mathrm{pt}~S^2}{\pi\sigma_S^2} \times \\ \exp\left[-\frac{N_\mathrm{pt} S^2}{2\sigma_S^2}\left[a^2+a_0^2-2aa_0\cos(2\phi-2\phi_0)\right]\right]a\,da\,d\phi
\end{multline}
where $\sigma_S$ is the uncertainty of $S$. Credibility intervals of $a$ and $\phi$ correspond to the intervals comprised between the 
minimum and maximum values of the parameter considered in the two dimensional 1 $\sigma$  contour plot (not shown).

\section{The spectral states of Cyg X-1}
\label{sec:asmstate}

We previously developed a method to classify the states of Cyg X-1 at any given time based on hardness and intensity measurements with the \emph{RXTE}/ASM and MAXI and on the hard X-ray flux obtained with the \emph{Swift}/BAT \citep{grinberg2013}. The ASM and MAXI data can be used to separate all three states (LHS, IS, HSS), while using BAT we can only separate the HSS, but not the LHS and IS from each other.

In microquasars in general and Cyg X-1 in particular, spectral transitions can occur on short time scales \citep[hours; e.g.,][]{boeck11}.  We thus performed the spectral classification for each individual ScW. This is a sufficiently small exposure over which to consider the source to be spectrally stable in the majority of the data. The IBIS 18--25\,keV light curve is shown in Fig.~\ref{fig:Ryle}, where the different symbols and colors represent the three different spectral states.

Where available, we used simultaneous ASM data for the state classification. ScWs with two simultaneous ASM measurements that are classified into different states are presumed to have ocurred during a state transition and excluded from the analysis. Most ScWs are, however, not strictly simultaneous with any ASM measurements.  To classify ScWs without simultaneous ASM measurements, we therefore used the closest ASM measurement within 6\,h before or after a given ScW. For the remaining pointings where no such ASM measurements exist, we used the same approach, first based on MAXI and then, if necessary, based on BAT.  As shown by \citet{grinberg2013}, the probability that a state is wrongly assigned using this method is 5\% within 6\,h of an ASM pointing for both the LHS and HSS, and $\sim$25\% for the IS.

Out of the 3302 IBIS ScWs, 1739 ($\sim$52.7\%) were taken during the LHS, 868 ($\sim$26.3\%) during the HSS and 316 ($\sim$9.6\%) during the IS. The remaining 379 (11.5\%) have an uncertain classification and are thus not considered in our analysis: 351 lack a simultaneous or quasi-simultaneous all sky monitor measurements that would allow a classification and 28 were caught during a state transition.

From MJD 55700 until MJD 56000, Cyg X-1 was highly variable and underwent several transitions on short timescales \citep[particularly striking in Fig.~1 of][]{grinberg2014}.  Inspection of some of the ScWs during this period shows that the source flux was very low in the ISGRI range. Since these ScWs were taken close to or in between state transitions, they were also removed from our analysis in order to limit the potential uncertainties they could introduce into the final data products. This filtering removed 32 additional ScWs.

The resulting IBIS exposure times are 2.05\,Ms, 1.21\,Ms, and 0.22\,Ms for the LHS, HSS, and IS, respectively. The IBIS 18--25\,keV and Ryle/AMI 15\,GHz light curves are shown in Fig.~\ref{fig:Ryle}.

\section{Results}

\subsection{Radio monitoring with the Ryle-AMI telescope}

\begin{figure*}
\epsscale{0.75}
\caption{The two plots represent the long term radio (15 GHz, upper panels) \integral/IBIS hard X-ray (18-25\,keV) light curves  (lower panels) of Cyg X-1. 
The different symbols and colors (online version) indicate the different states of the source according to the classification 
presented by \citet{grinberg2013}. The upper plot covers the period MJD 52760--54250, when the radio monitoring was  made 
with the older implementation of the radio telescope (known as the Ryle telescope). The lower one covers MJD 54500--56800 
and shows the radio monitoring made after the upgrade (AMI telescope). Note the different vertical scales 
for the radio fluxes between the two plots. }
\plotone{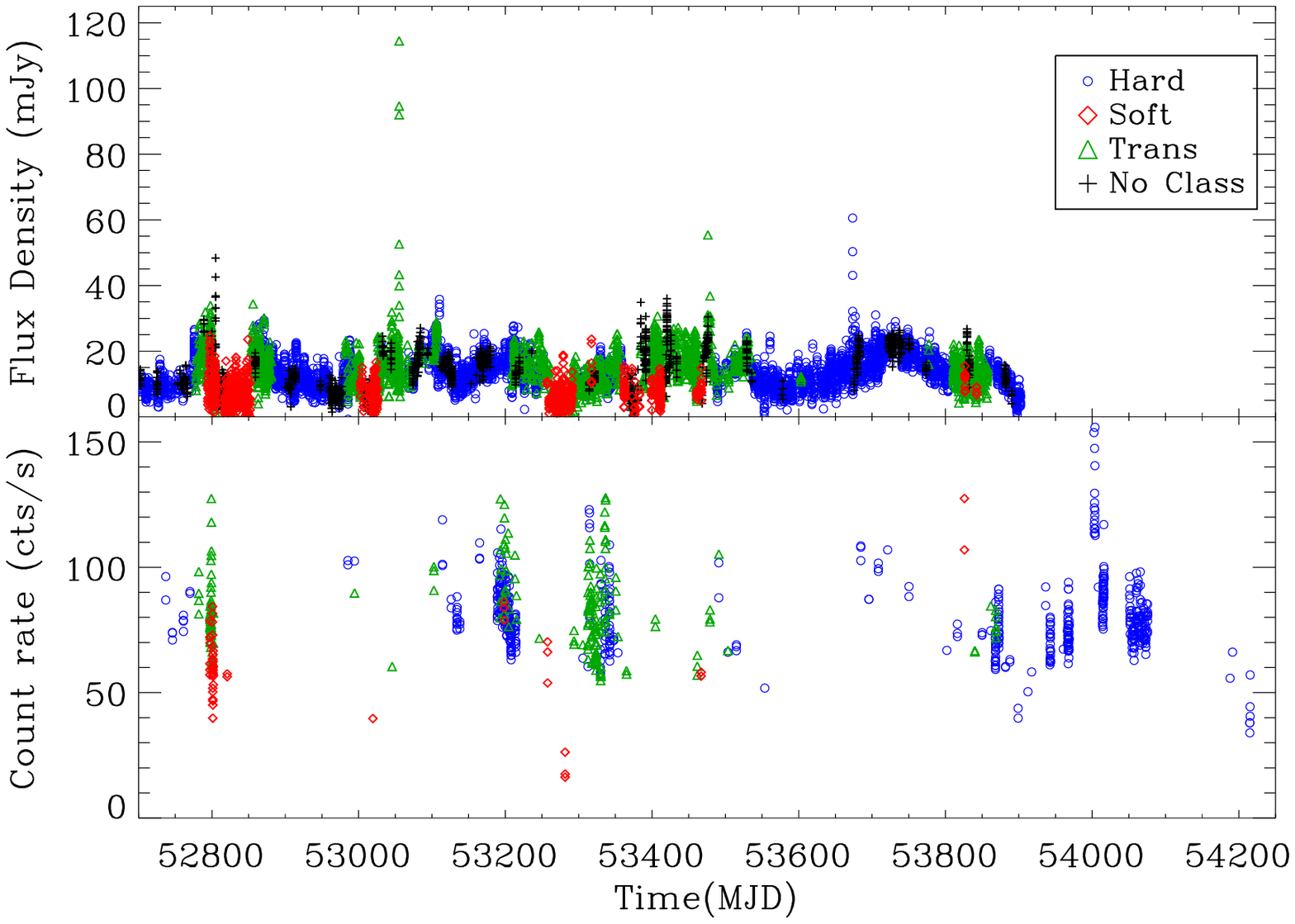} \\
\plotone{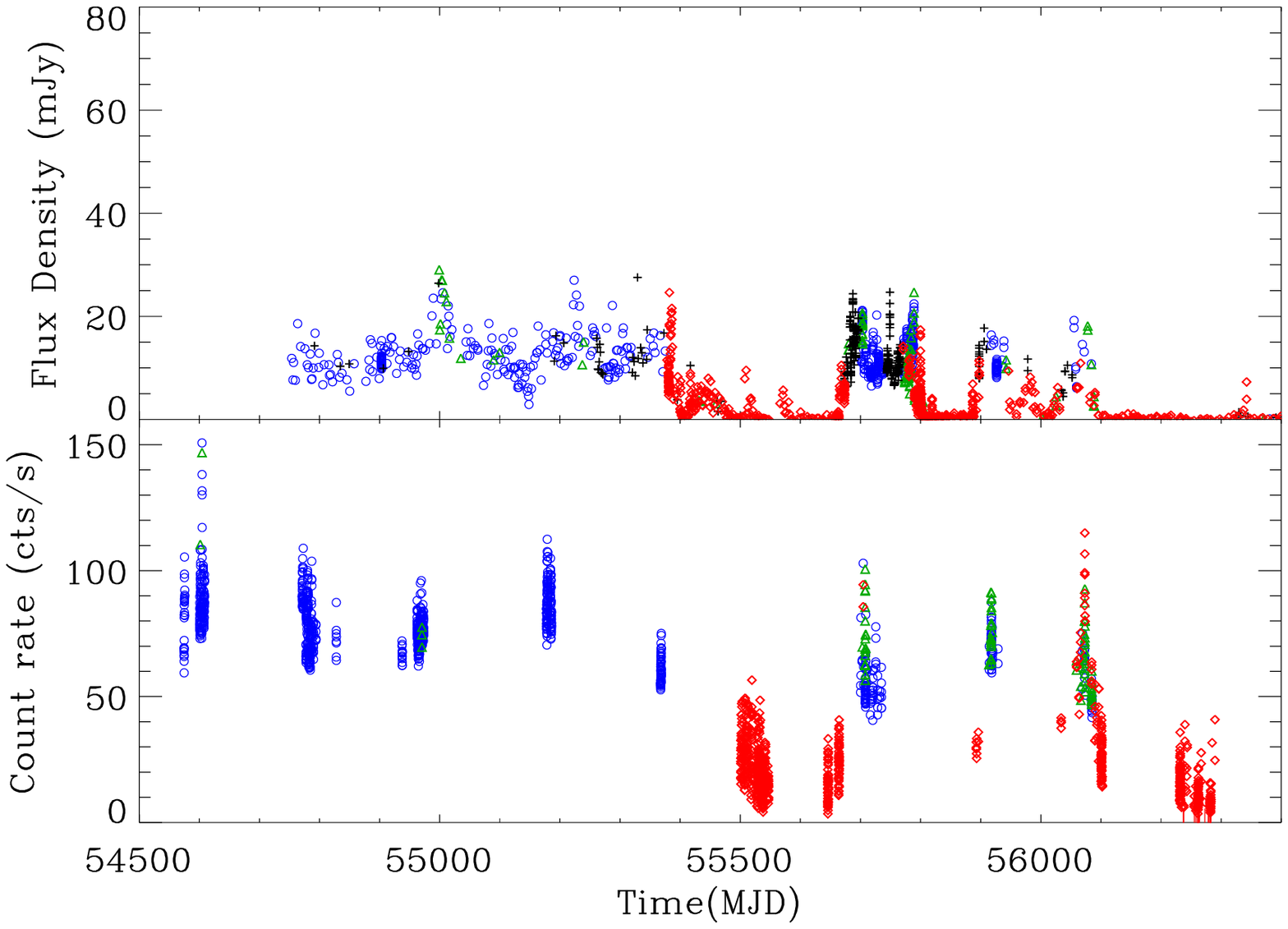}
\label{fig:Ryle}
\end{figure*}

The 2002--2012 15\,GHz light curve classified utilizing the same method as for the \integral\ data is shown in Fig.~\ref{fig:Ryle}. As known from previous studies of microquasars, including Cyg X-1, the LHS and the IS show a high level of radio activity, while the radio flux is at a level compatible with or very close to zero during the soft state \citep[e.g.,][]{Corbel03,Gallo03}.

Cyg X-1 shows a high level of radio variability with a mean flux $\langle F_{\mathrm{15\,GHz}}\rangle\sim12$ mJy and flares. The most prominent flare occurred on MJD 53055 (2004 February 20) and reached a flux of 114\,mJy\footnote{This flare was studied in detail by \citet{fender06}, although the time axis of their Fig.~4 is wrongly associated with MJD 55049.2--55049.6}. Flares occur frequently (Fig.~\ref{fig:Ryle}), and are connected with the X-ray behavior of the source \citep{fender06,wilms07}.  The main flares seem to mostly coincide with the IS. The radio behavior in the LHS is usually steadier (although there is, e.g., a flare at MJD~53700, Fig.~\ref{fig:Ryle}).

We obtained state resolved mean radio fluxes of $\langle F_{\mathrm{15\,GHz, LHS}} \rangle=13.5$\,mJy, $\langle F_{\mathrm{15\,GHz,IS}} \rangle=15.4$\,mJy, and $\langle F_{\mathrm{15\,GHz, HSS}} \rangle=4.6$\,mJy. Given the typical rms uncertainty of $\mathrm{rms(5min)}=3.0$\,mJy \citep[e.g.][]{pooley97,rodrigue08_1915a}, the mean radio flux in the HSS is not detected at the $3\sigma$ level.  The few clear radio detections in this state (Fig.~\ref{fig:Ryle}) do not correspond to a steady level of emission that would indicate a compact radio core \citep[see also][]{fender99}. Such detections preferably occur after radio flares and thus likely represent the relic emission of previously ejected material \citep[e.g.][]{Corbel04b}.
 
\subsection{State resolved spectral analysis}

We analyzed the state resolved energy spectra of each instrument using \texttt{XSPEC v.12.8.0} \citep{Arnaud1996}. For JEM-X we considered the 10--20\,keV range.\footnote{Although the JEM-X units are well calibrated above $\sim$5\,keV, we chose to ignore the lower energy bins. Due to the possible influence of a disk and presence of iron fluorescence line at 6.4\,keV they would add uncertainties and degeneracies to the models that the low energy resolution does not allow us to constrain.} and 20--400\,keV for the ISGRI data. We used Compton data between 300\,keV and 1 or 2\,MeV, depending on their statistical quality.  We started by fitting the spectra in the 10--400\,keV range since the contribution of the 1\,MeV power-law tail is supposed to be negligible here (LRW11). When an appropriate model was found for this restricted range, we added the 0.3--2\,MeV Compton spectra and repeated the spectral analysis.

In order to find the most appropriate models describing the 10--400\,keV spectra for all three spectral states, we proceeded in an incremental way.  We started with a simple power-law model and added spectral model components until an acceptable fit according to $\chi^2$-statistics was found. The phenomenological models were then replaced by more physical Comptonization models.  Since the $\geq 10$\,keV data are not affected by the absorption column density of $\mathrm{N}_\mathrm{H} \sim$a few $10^{22}\,\mathrm{cm}^{-2}$ \citep[e.g.,][]{boeck11,Grinberg_2015a}, no modeling of the foreground absorption is necessary.  Cross-calibration uncertainties and source variability during different exposure times are taken into account by a normalization constant. The ISGRI constant was frozen to one and the others left free to vary independently. We required that the values of the free constants were in the range $[0.85,1.15]$.  The bestfit parameters obtained with the best phenomenological and with the Comptonization models are reported in Table~\ref{tab:fits1} for all three spectral states.

\subsubsection{Hard state}
\label{sec:hardstate}
\begin{figure*}[htpb]
\caption{{\bf{Left:}} $\chi^2$ residuals to the hard state 10--400\,keV spectra. From a) to d) simple power-law, power-law with 
 exponential cut-off, Comptonization,   Comptonization convolved 
by a reflection model. {\bf{Right:}} 10--500\,keV $\nu\,-F_\nu$ (JEM-X+ISGRI) spectrum with the best model (Comptonization convolved by a reflection model) superimposed as
a continuous line. For the sake of clarity only one JEM X spectrum is represented. }
\epsscale{1}
\plotone{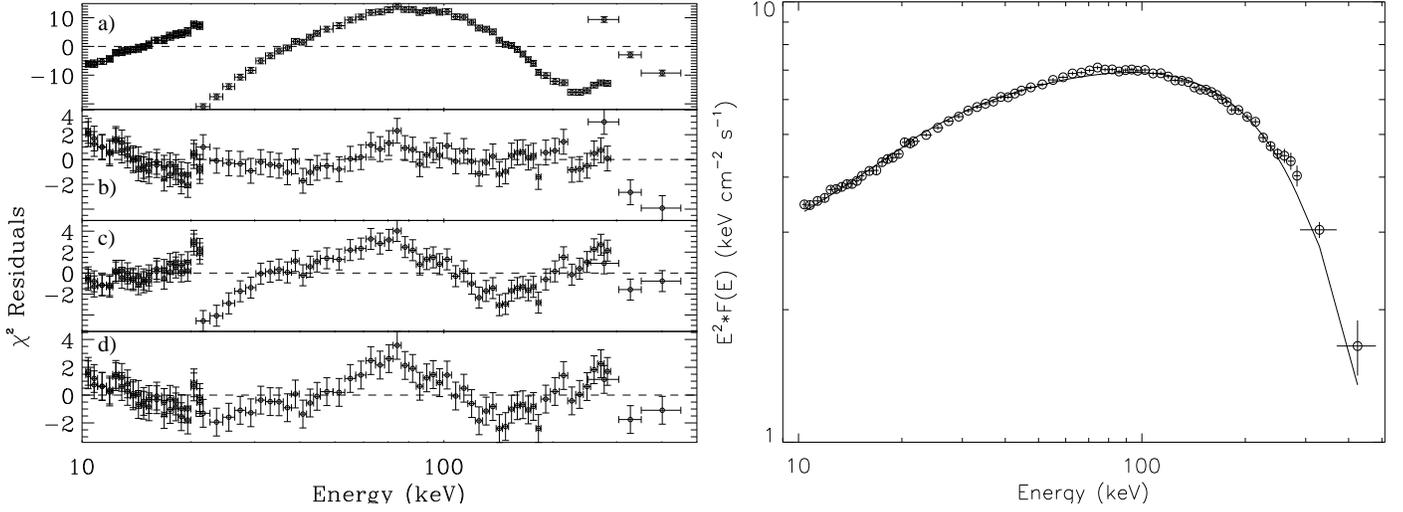}
 \label{fig:reshard}
\end{figure*}

A power-law (Fig.~\ref{fig:reshard}a) gives a very poor representation of the 10--400\,keV data with a reduced $\chi^2= \chisq \sim 70$ for 87 degrees of freedom (dof). The broad band spectrum clearly departs from a straight line, which indicates a curved spectrum. A power-law modified by a high energy cut-off (\texttt{highecut*power} in \texttt{XSPEC}) provides a highly significant improvement with $\chisq=0.89$ for 85 dof (Fig.~\ref{fig:reshard}b). We note, however, that the value of the high energy cut-off is unconstrained (Table~\ref{tab:fits1}), and deviations are seen at high energies.

As a cut-off power-law in this energy range  is usually assumed to mimic the spectral shape produced by an inverse Compton process, we replaced the phenomenological model by a more physical model \texttt{comptt} \citep{titarchuk94, Titarchuk_Lyubarskij_1995a,
  Titarchuk_Hua_1995a}. Due to our comparably high lower energy bound, the temperature of the seed photons could not be constrained. We fix it to a typical LHS value of 0.2\,keV \citep{Wilms_2006a}.  Although the high energy part is better represented, the fit is statistically unacceptable ($\chisq=2.9$ for 88 dof, Fig.~\ref{fig:reshard}c) and significant residuals are visible in 10--80\,keV range. In this range, hard X-ray photons can undergo reflection on the accretion disk. This effect is usually seen as an extra curvature, or bump in the 10--100\,keV range in the spectra of microquasars and AGNs \citep[][and references therein]{Garcia_2014b}. We thus included a simple reflection model (\texttt{reflect}) convolved with the Comptonization spectrum and fixed the inclination angle to $30^\circ$. The residuals are much smaller, yielding an acceptable fit ($\chisq=1.67$ for 87 dof)\footnote{We have followed the recommendations of the IBIS user manual regarding the level of systematics added to the IBIS spectral points (http://isdc.unige.ch/integral/download/osa/doc/10.0/osa\_um\_ibis/node74.html). Systematic errors of 1\% seem to underestimate the real uncertainties when dealing with (large) data sets that encompass several calibration periods. We have tested this notion by refitting the spectra with 1.5\% and 2\% systematics and, as expected, obtain \chisq\ much closer to 1. Since the values of the spectral parameters, and thus the conclusions presented here, do not change significantly, we decided to present the spectral fits using the recommended 1\% for systematics.}.  The normalization constants are 0.99, 1.06, and 0.98 for JEM X-1, JEM X-2, and the Compton mode respectively. The best fit results of this model are listed in Table~\ref{tab:fits1}, while the $\nu\,F\nu$ broad band 10--400\,keV spectrum is shown in the rightmost panel of Fig~\ref{fig:reshard}.

\begin{table*}
\caption{Spectral parameters of fit to the 10--400\,keV spectra. Errors and limits on the spectral parameters are given at the $90\%$ confidence 
level, while the errors on the fluxes are at the $68\%$ level.}
\begin{tabular}{ccccccc}
\hline
\multicolumn{7}{c}{{\tt{reflect*highecut(powerlaw)}}}\\
\hline
State &   $\Gamma$ & E$_{\mathrm{cut}}$ & E$_{\mathrm{fold}}$ & \multicolumn{3}{c}{Fluxes$^\ddagger$} \\
          &                        &         (keV)                  &      (keV)                       &10--20\,keV  & 20--200\,keV& 200--400\,keV\\
\hline
LHS &     $1.43\pm0.01$    &        $\le 12$ & $155\pm4$  &     $4.48 \pm0.02$     & $22.30 \pm0.03$ & $3.56\pm0.03$\\        
IS &  $1.87_{-0.03}^{+0.02}$ & $56_{-6}^{+4}$ & $198\pm8$ &  $5.10\pm0.03$ & $18.47\pm0.04$ & $2.52\pm0.05$\\
HSS$^\dagger$ & $2.447\pm0.007$ &$130_{-16}^{+11}$ & $198_{-59}^{+135}$ &$1.83\pm0.01$ & $3.33\pm0.01$ & $0.18\pm0.03$\\

\hline
\multicolumn{7}{c}{{\tt{reflect*comptt}}}\\
\hline
State & $\Omega/2\pi$ &  kT$_{e}$ & $\tau$ & \multicolumn{3}{c}{Fluxes$^\ddagger$} \\
                &  & (keV)       &           & 10--20\,keV  & 20--200\,keV& 200--400\,keV\\
\hline
LHS & $0.13\pm0.02$  &  $59.4_{-1.2}^{+1.3}$ & $1.06\pm0.03$ & $4.37 \pm0.03$ & $22.60\pm0.03$ & $3.70\pm0.03$\\
IS & $0.04\pm0.03$  &  $54.4_{-2.8}^{+3.6}$ & $0.82\pm0.06$ & $5.30 \pm0.03$ & $18.42\pm0.06$ & $1.93\pm0.07$\\
HSS & $0.36\pm0.03$ & $279\pm15$ & $<0.013$ & $1.84 \pm0.09$& $3.7\pm0.6$ & $0.36\pm0.01$\\
\end{tabular}
\begin{list}{}{}
\item[$^\dagger$]A reflection component with $\Omega/2\pi$=0.46$\pm0.04$ was included for a good spectral fit to be obtained
\item[$^\ddagger$]In units of $10^{-9}$~\ergcms
\end{list}
\label{tab:fits1}
\end{table*}

\begin{figure*}[htpb]
\epsscale{1}
\caption{Unfolded 10\,keV--2 MeV \integral\ spectra of Cyg X-1 in the three spectral states. The LHS spectrum is in blue circle, the IS in green diamonds and the HSS 
one in red triangles (colored points are available in the online version).}
\plotone{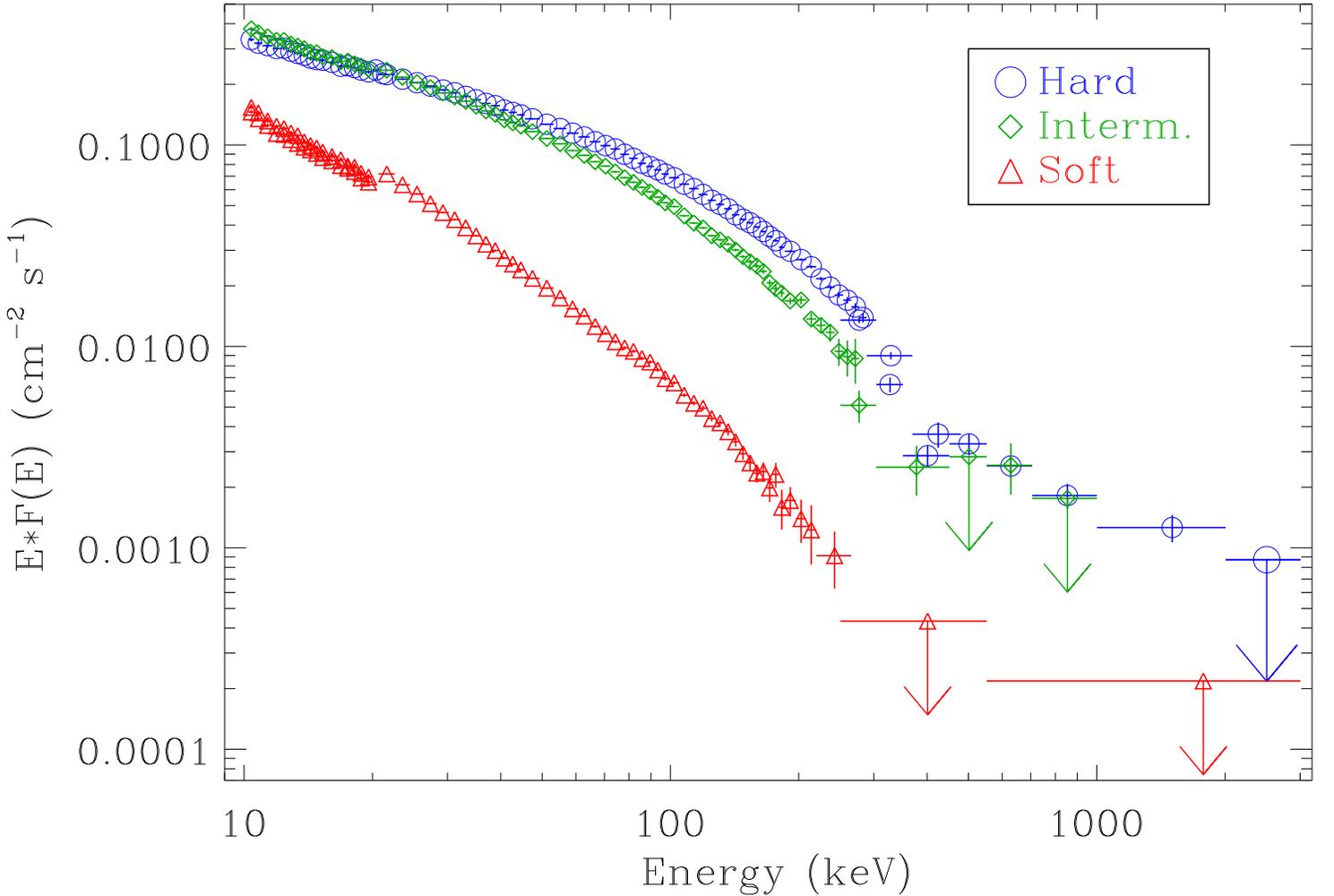}
 \label{fig:Allspec}
\end{figure*}

Next, we added the $\geq$400\,keV Compton mode spectrum (Fig.~\ref{fig:Allspec}) to our data. These points are significant at the $8.2\sigma$, $7.8\sigma$, $6.9\sigma$, and $6.5\sigma$ levels in the 451.4--551.9\,keV, 551.9--706\,keV, 706--1000.8\,keV, and $\sim$1--2\,MeV energy bands. A significant deviation from the previous best fit model is clearly visible above 400\,keV. An additional power-law improves the fit significantly ($\chisq=1.4$ for 90 dof). The photon index of this additional power-law is hard ($\Gamma=1.1_{-0.4}^{+0.3}$) and marginally compatible with the values reported earlier ($1.6\pm0.2$, LRW11).

We model the reflection component using several distinct approaches, assuming 1) that the power-law component undergoes the same reflection as the Comptonization component (\texttt{reflect*(power+comptt)}), 2) that both components undergo reflection but with different solid angles $\Omega/2\pi$ (\texttt{reflect$_{1}$*(power)+reflect$_{2}$*(comptt)}), and 3) that the power-law is not reflected (\texttt{power+reflect*(comptt)}). In model~2, the value for the reflection of the power-law component is unconstrained. Model~3 results in a better \chisq, but the value of the power-law parameters are physically not acceptable. The normalization is too high and $\Gamma$ too soft to reproduce the Compton data well.  We therefore discard model~3, and consider model~1 as the most valid one. The value of the fit statistics is, however, still high with $\chisq=1.8$ for 89 dof. The residuals show deviations in the $\sim50$--200\,keV range that may be due to the accumulation of a large set of ScWs taken at different calibration epochs and at different off-axis and roll angles of Cyg X-1.

In the final fit to the 10\,keV--2\,MeV broad band spectrum the parameters of the Comptonized component are $kT_e=53\pm2$\,keV and $\tau=1.15\pm0.04$, the reflection fraction is unchanged, and the photon index of the hard tail is $\Gamma=1.4_{-0.3}^{+0.2}$, i.e., essentially compatible with the value reported in LRW11. The 0.4--1 MeV flux of the hard tail ($F_{\mathrm{pow,0.4-1~MeV}}=1.9\times10^{-9}\,\ergcms$) accounts for 86\% of the total 0.4--1 MeV flux.

\subsubsection{Intermediate state} 
A simple power-law gives a poor representation of the IS spectrum ($\chisq\sim28$ for 82 dof).  The inclusion of a high energy cut-off again greatly improves the fit ($\chisq=1.35$, 80 dof), but the normalization constant of the Compton spectrum ($C_{\mathrm{Compton}}$) is outside the range that is consistent with the flux calibration of the instrument.  Replacing the phenomenological model by a \texttt{comptt} model, with the seed photon temperature fixed at 0.8\,keV\footnote{The choice of a higher seed photon temperature in the IS does not influence the $>10$~keV spectrum and was only motivated by the desire to be consistent with the expectation of a hotter disk in the IS.} leads to a good fit ($\chisq=1.1$ for 81 dof) and a physical value of $C_{\mathrm{Compton}}$.  Although the statistics are not as good as in the LHS due to the much shorter total exposure, and the residuals in the 10--100\,keV region are rather acceptable, we included a reflection component to be consistent with the LHS modeling. This approach improved the fit to $\chisq=0.97$ for 79 dof, which indicates a chance improvement of $\sim1.2\%$ according to an F-test (the reflection component has a significance lower than $3\sigma$). The reflection fraction is very low and poorly constrained ($\Omega/2\pi=0.04\pm0.03$). As the other parameters are not significantly affected, we still report the results obtained with the latter model in Table~\ref{tab:fits1} despite this weak evidence for reflection.

We then added the $\geq 400$\,keV spectral points of the Compton mode spectrum (Fig.~\ref{fig:Allspec}). The previous model (without reflection) leads to a good representation ($\chisq=1.2$ for 83 dof), even if a deviation at high energies is present.  Above 400\,keV, however, only the 551.9--706\,keV Compton mode spectral point is significant at $\geq 3\sigma$. Adding a power-law to the data leads to $\chisq=1.04$ (81 dof), which corresponds to a chance improvement of 0.1\%. Not surprisingly, the power-law photon index $\Gamma$ is poorly constrained ($-0.4\leq\Gamma\leq1.8$) and the normalization constant for the Compton mode spectral points $C_{\mathrm{Compton}}$ tends to a very low value if not forced to be above 0.85. Fixing $\Gamma$ at 1.6 (LRW11, JRC12)
 and  setting $C_{\mathrm{compton}}$ to 0.90, the best value found in the LHS fits, allows us to estimate a $3\sigma$ upper limit 
for the 0.4--1\,MeV flux of $1.5\times 10^{-9}\,\ergcms$.

\subsubsection{Soft State}
In the soft state the soft X-ray spectrum below 10\,keV is dominated by a $\sim$1\,keV accretion disk. This disk, however, does not influence the $>$10\,keV spectrum studied here.  A single power-law gives a poor representation of the broad band spectrum ($\chisq=6.0$ for 77 dof). The JEM-X and ISGRI spectra are particularly discrepant and a possible hint for a high energy roll-over is seen in the residuals. A cut-off power-law greatly improves the fit statistics ($\chisq=1.68$ for 75 dof), but the modeling is still not satisfactory and the JEM-X normalization constants are inconsistent with the detector calibration uncertainty. Residuals in the 15--50\,keV range suggest that reflection on the accretion disk may occur.  Adding the \texttt{reflect} model improves the fit to $\chisq=1.24$ (74 dof). The normalization constant for the JEM-X1 detector has to be forced to be above 0.85 as it would naturally converge to about 0.8. Alternatively describing the data with a simple reflected power-law without a cutoff results in  $\chisq=1.88$ (76 dof) and shows that the cut-off is genuinely present. The chance improvement from a reflected power-law spectrum to a reflected cut-off power-law spectrum is about $4\times10^{-8}$ according to an F-test. Replacing the phenomenological model with a Comptonization continuum that is modified by reflection yields $\chisq=1.69$ (75 dof, Table~\ref{tab:fits1}).

The $>$400\,keV data were then added. Since the Compton mode spectrum has a very low statistical quality and in order not to increase the number of degeneracies, we fixed its constant to a value of one. The \texttt{reflect*comptt} model is still statistically acceptable, and only the highest spectral bin (550--2000\,keV) indicates an excess of source photons compared to the model ($\chisq=1.76$, 77 dof). Adding a powerlaw component to the overall spectrum improves the modeling of the spectrum to $\chisq=1.69$ (75 dof), but if left free to vary $\Gamma$ is completely unconstrained. We fix it at 1.6 ($\chisq=1.70$, 76 dof, 0.11 chance probability according to the F-test) and derive a $3\sigma$ upper limit of $0.93\times10^{-9}\,\ergcms$ for the 0.4--1~MeV flux.

\subsection{State resolved polarization analysis}
\label{sec:res-polar}
Polarigrams were obtained for each of the three spectral states.  The low statistics of the IS (the source significance in the Compton mode is $5.7 \sigma$) does not permit us to constrain the 250--3000\,keV polarization fraction.

In the HSS, the detection significances of Cyg X-1 with the Compton mode are $4.2\sigma$ and $7.7\sigma$ in the 300--400\,keV and 400--2000\,keV bands respectively. A weighted least 
square fit procedure was used to fit the polarigrams. A constant represents the 400--2000~keV polarigram rather well ($\chisq=1.6$, 5 dof, not shown). 
The $1\sigma$ upper limit of the 400--2000\,keV polarization fraction in the HSS is $\sim$70\%.

\begin{figure*}[!htbp]
\caption{Polarigrams obtained in the LHS.   {\bf{Left: }} 300--400\,keV. {\bf{Right:}} 400--2000\,keV. }
\epsscale{1}
\plottwo{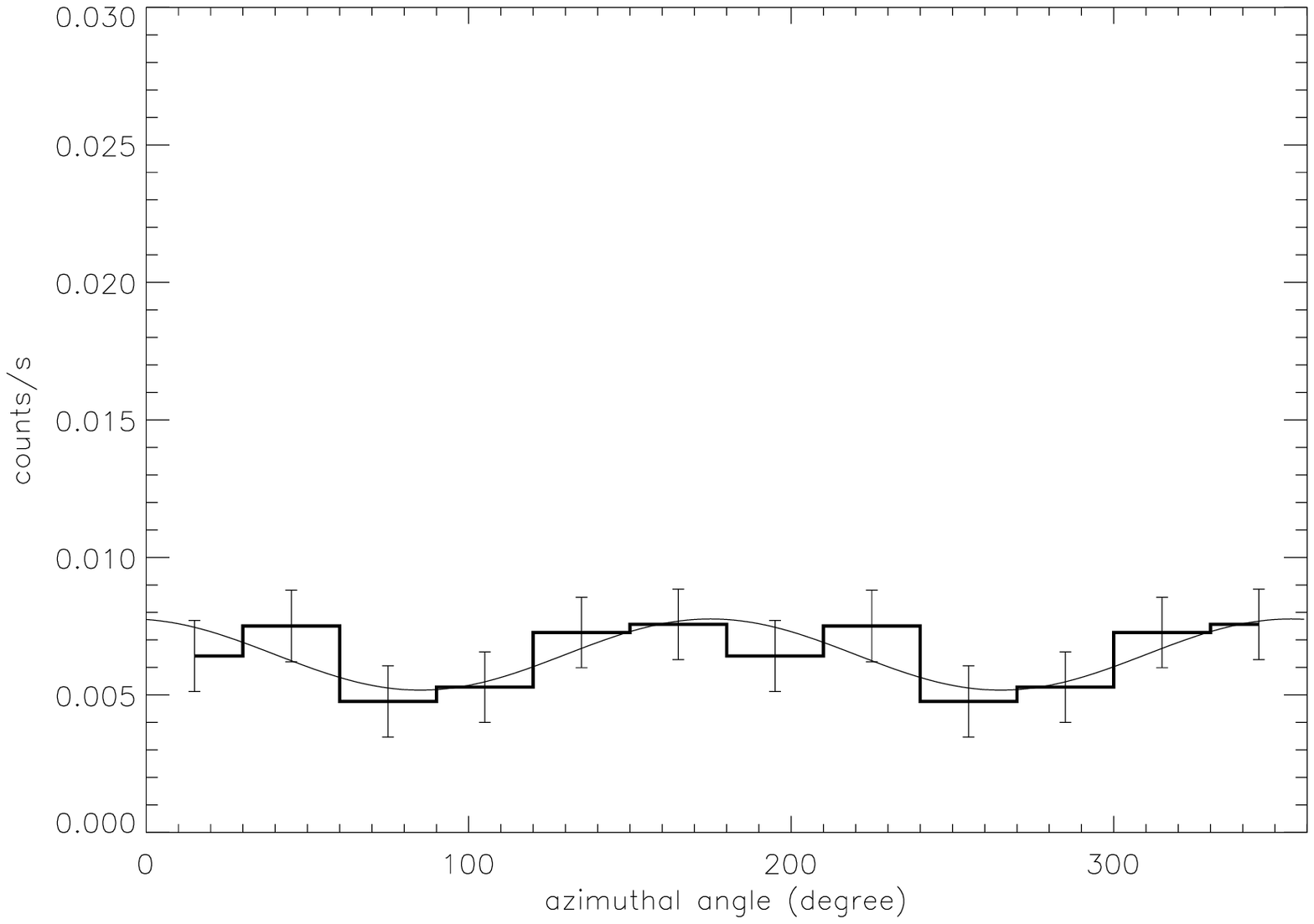}{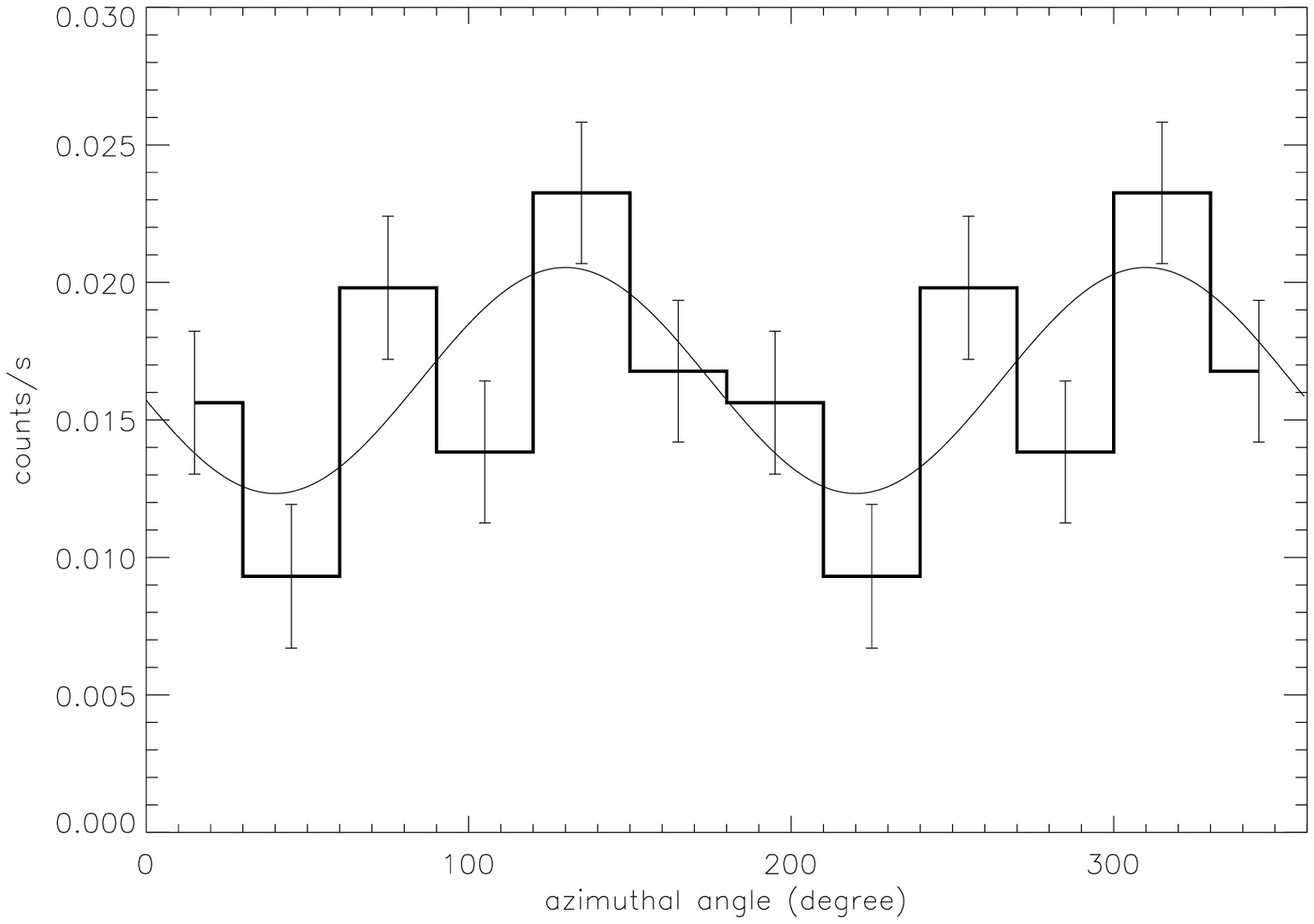}
\label{fig:pola}
\end{figure*}

Figure~\ref{fig:pola} shows the LHS polarigrams in two energy ranges: 300--400\,keV and 400--2000\,keV. Note that the data selection 
described in Sec.~\ref{sec:asmstate} is at the origin of the different count rates between the polarigrams shown here and those in Fig. 2 or LRW11. 
This effect is more obvious at the highest energies where the different spectral slopes have the largest impact (Fig. ~\ref{fig:Allspec}).  In the lower energy range, the Compton 
mode detection significance of Cyg X-1 is $12.3\sigma$.  The polarigram can be described by a constant ($\chi^2_\nu=0.87$, 5 dof). We estimate an 
upper limit for the polarization fraction of $\Pi_{300-400~{\mathrm{keV, LHS}}}=22\%$ in this energy range.  The 400--2000\,keV energy range 
polarigram shows clear deviation from a constant and a constant poorly represents the data ($\chi^2_\nu=3.5$, 5 dof). The pattern and fit indicate 
that the photons scattered in ISGRI are not evenly distributed in angle, but that the scattering has a preferential angle as is expected from a polarized 
signal (\S2.2). We describe the distribution of the polarigram with the expression of Eq.~\ref{eq:azimuth} and obtain the values of the parameters $a_{0}$ 
and $\phi_{0}$ with a least square technique.  The fit is better than that to a constant, although not formally good ($\chi^2_\nu=2.0$, 3 dof) due tothe low 
value of the periodogram at azimuthal angle $100^\circ$ (Fig.~\ref{fig:pola}).  Assuming that the non constancy of the polarigram is an evidence for a real 
polarized signal \citep[an assumption also consistent with the independent detections of polarized emission in Cygnus X-1 with SPI; ][]{jourdain12}, 
we estimate that the polarization fraction is $\gtrsim 10\%$ at the 95\% level, and greater than a few \% at the 99.7\% level. In other words we obtain a 
significance higher than $3\sigma$ for a polarized 400--2000\,keV LHS emission.  We determine a polarization fraction and a credibility interval 
of  $\Pi_{\mathrm{400-2000\,keV, LHS}}=75\pm 32 \%$ at position angle $\mathrm{PA}_{\mathrm{Cyg X-1, LHS}} = 40.0^\circ \pm 14.3^\circ$.  The values 
of the polarization fraction and angle are consistent with those found by JRC12. We note that an error of 180$^\circ$ in the angle convention in 
LRW11 led to $\mathrm{PA} = 140^\circ$, which after correction becomes $40^\circ $, in agreement with our state resolved result (see also JRC12). 
Note that the good agreement of our results with those obtained with SPI in the case of Cyg X-1, but also the fact that IBIS also detected polarization in the high energy emission of 
 the Crab nebula and pulsar \citep{forot08} at values compatible with those obtained with SPI \citep{dean08}, increase the confidence in the 
 reality and validity of our results.

\section{Discussion and conclusions}

\subsection{The  $\geq 400$~keV hard tail}
\begin{figure*}[!ht]
\caption{{\bf{Left:}} Comparison of the \integral/IBIS and {\it{CGRO}}/Comptel high energy spectra. {\bf{Right:}} Comparison of the 
\integral/Compton spectra accumulated over two different epochs.}
\epsscale{1}
\plottwo{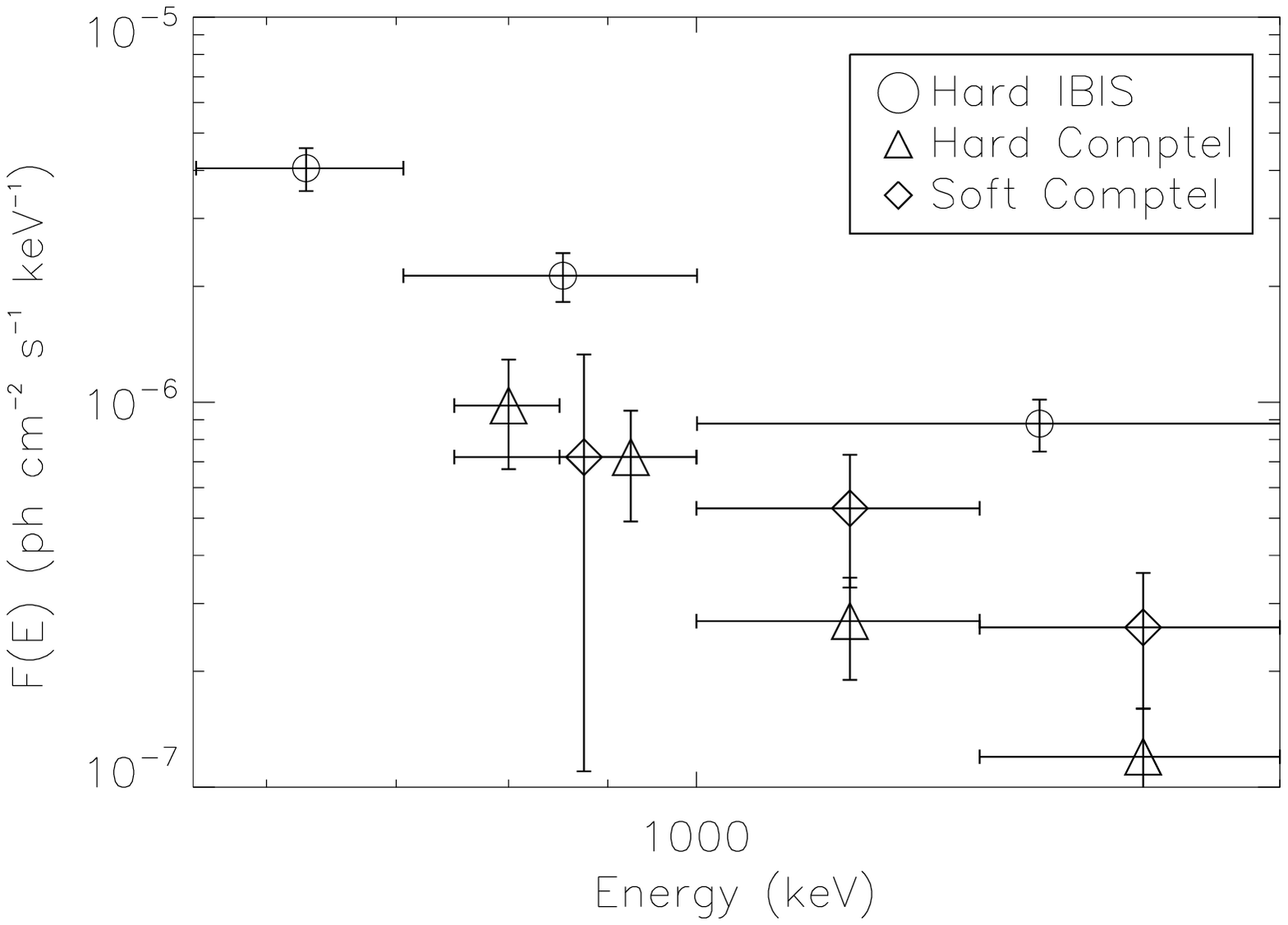}{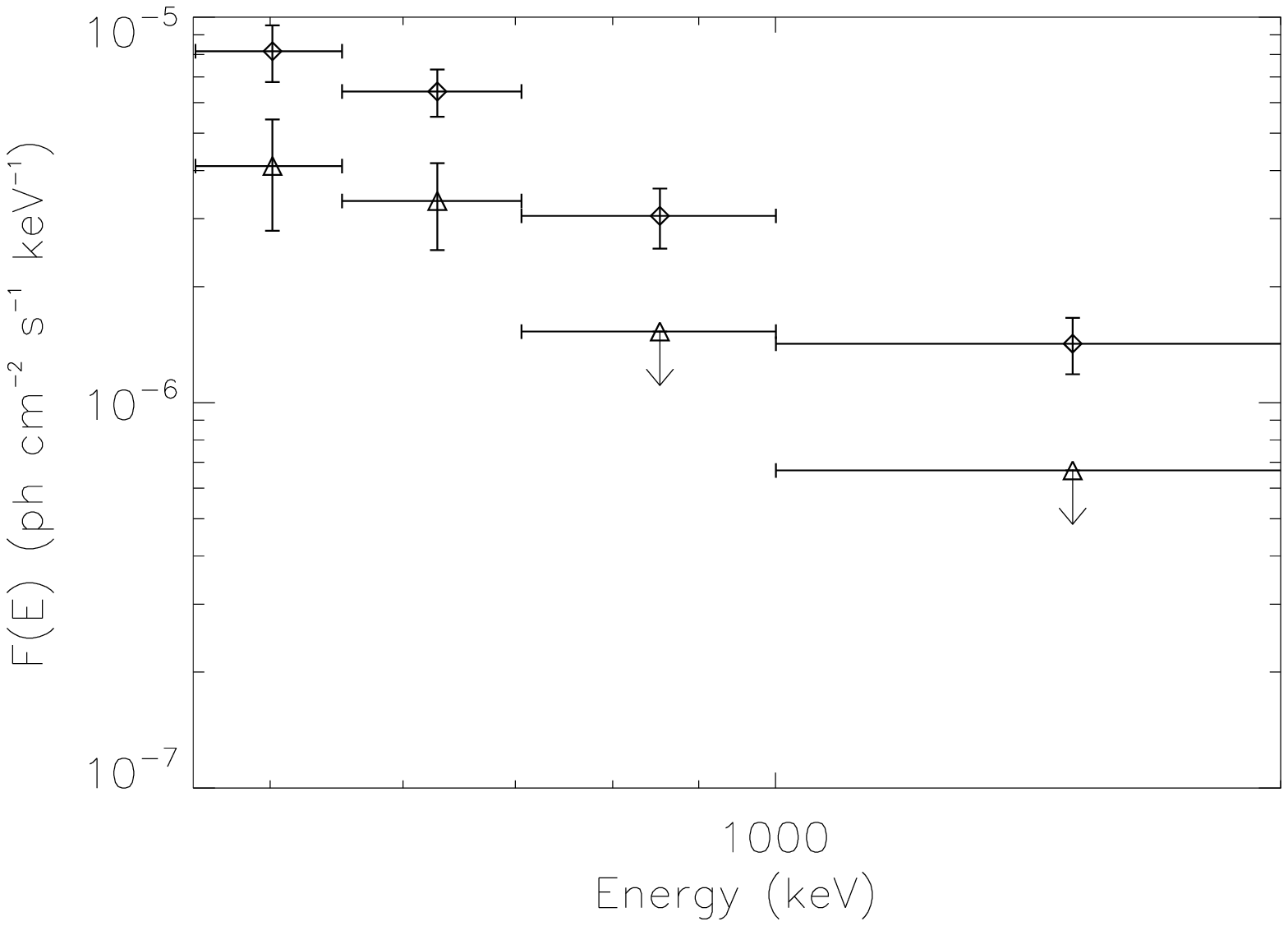}
\label{fig:comptel}
\end{figure*}

Our analysis shows that the $>400$\,keV spectral properties of Cyg X-1 depend on the spectral state, as already suggested previously \citep[e.g.,][]{mcconnell02}. We detect a strong hard tail in the LHS, but obtain only upper limits in the IS and HSS. In Fig.~\ref{fig:comptel} (left), we show a comparison of our LHS 0.75--1\,MeV spectrum with those obtained with \textit{CGRO}/COMPTEL during the June 1996 HSS and during an extended LHS observed with this instrument \citep{mcconnell02,mcconnell00}. The IBIS hard tail is compatible in terms of fluxes to within $2\sigma$ to those seen with COMPTEL and also with the \integral/SPI hard tail reported in JRC12 (see their Fig.~5).  It is worth noting that Cyg X-1 is a notoriously highly variable source and flux evolution between different epochs is expected. To illustrate this behavior, we study data taken during two main periods, the covering March 2003 to December 2007 (MJD 52722--54460) and March 2008 to December 2009 (MJD 54530--55196), respectively. These intervals correspond to the analysis of LRW11. The two Compton spectra are shown in Fig.~\ref{fig:comptel} (right).  It is obvious that the flux of the hard tail varied between the two epochs and an evolution of the slope may also be visible.  We note that these data are not separated by states, but as they cover the period before December 2009, they are dominated by the LHS (e.g., Fig.~\ref{fig:Ryle}). We therefore conclude that, even in the same state, the hard tail shows luminosity variations.

A major difference to \citet{mcconnell02} is the non-detection of a hard tail in the HSS.  \citet{mcconnell02} remark that between two epochs of HSS the hard tail either has a higher flux than in their averaged LHS or is not detected.  \citet{malyshev2013} also note that the 1996 HSS hard tail comes from a single occurrence of this state and may not reflect a typical behavior.  While detection of a hard tail during the LHS, even if it shows variations, indicates an underlying process that is rather stable over long period of times, the apparent transience of the HSS hard tail may indicate a different origin for this feature. For example, in Cyg~X-3 the very high energy (Fermi) flares are associated with transitions to the ultra soft state \citep{abdo2009}. While this is similar to what is seen in Cyg~X-1, all Fermi $\gamma$-ray flares of Cyg X-1 were reported during LHS \citep{bodaghee13,malyshev2013}.  While the lack of simultaneous very high energy GeV-TeV data, and/or polarimetric studies of the Comptel HSS hard tail prevent any further conclusion on the origin of this component, we conclude that the HSS hard tail and the LHS hard tail could be of different origin.

\subsection{Polarization of the hard tail and its possible origin}
A clear polarized signal is detected from Cyg~X-1 with both IBIS and SPI while accumulating all data regardless of the spectral state of the 
source (LRW11, JRC12).  Here we separated the data into different spectral states and studied the state dependent high energy polarization 
properties. The HSS and 300--400\,keV polarigram can be described by a constant, indicating no or very little levels of polarization. The 
400--2000\,keV LHS polarigram shows a large deviation from a constant, which indicates that the Compton scattering from ISGRI is not an 
evenly circular distribution on PICSiT.  We interpret this as evidence for the presence of polarization in this energy range, even if the polarigram 
shows some deviations from the theoretically expected curve (Fig.~\ref{fig:pola}). In the 300--400\,keV range, no evidence for polarization is
 found. This is consistent with the results obtained with SPI (JRC12) and shows that the polarization fraction is strongly energy dependent. 
 Assuming a non-null level of polarization, a reasonable assumption given the above arguments, we estimate that the detection of polarized 
 emission is significant at higher than $3\sigma$.  Given the larger amount of data and the separation into different spectral states, 
 one could have expected to obtain more robust results, and more constrained polarization parameters compared to LRW11 and JCR12, 
 while the uncertainties we obtained are, at best, of the same order. The reason for this lies in the behavior of Cyg~X-1. While the earlier 
 studies did not perform a state dependent analysis, our state classification reveals that both studies were dominated by data taken during 
 the LHS and the IS (see also Fig.~\ref{fig:Ryle}). In fact, compared to LRW11 the number of ScWs measured during the LHS increased 
 only by 13\%, and 75\% of the IS data comes from the period considered in these earlier studies, i.e., before $\sim$MJD~55200. On the 
 other hand, about 94\% of the HSS data are new. It is therefore not surprising that our hard state result is consistent with the earlier results, 
 as these were fully dominated by the (polarized) hard state data. The exposure of the less polarized IS data was not large enough to 
 significantly ``dilute'' the polarization. Our result clearly shows that larger amounts of data in each of the states are needed to refine the 
 constrain on the polarization properties and their relation to the spectral states.

Our LHS spectral analysis shows that the 10--1000\,keV spectrum can be decomposed into several spectral components. A (reflected) thermal Comptonization component in the range $\sim10$--400\,keV, and a power-law tail dominating above $\sim$400\,keV (\S\ref{sec:hardstate}).  The origin of the seed photons for the Compton component may either be an accretion disk or synchrotron photons from a compact jet undergoing synchrotron self-Comptonization (SSC).  As discussed elsewhere \citep[e.g.,][]{laurent11}, due to the large number of scatterings the Comptonization spectrum from a medium with an optical depth $\tau\ge 1$ is not expected to show intrinsic polarization, especially with such a high fraction as the one we detect here. Multiple Compton scattering, as expected in such a medium, will ``wash out'' the polarization of the incident photons even if the seed photons are polarized (e.g.,  jet photons).  It is therefore not surprising that the LHS $\lesssim$400\,keV component has a very low (or zero) polarization fraction \citep[see also][]{russell13}.

The origin of the hard-MeV tail is much less clear and is highly model-dependent.  Two main families of models can explain the (co-)existence of cut-off power-law and/or pure power-law like emissions in the energy spectra of BHBs.  The first is based on the presence of an hybrid thermal-non thermal population of electron in a ``corona'', and hybrid Comptonization have been successfully applied to the spectra of, e.g., GRO J1655$-$40 or GX 339$-$4 \citep[e.g.,][]{caballerog09, joinet07}, and even to $\sim1$--10000~keV spectra of Cyg X-1, in both the HSS and LHS \citep{mcconnell02}.  A double thermal component has been used to represent the 20--1000\,keV SPI data of 1E 1740$-$2947 well \citep{bouchet09}.

The second family of models essentially contains the same radiation processes but is based on the presence of a jet emitting through direct synchrotron emission, thermal Comptonization of the soft disk photons on the base of the jet, and/or SSC of the synchrotron photons by the jet's electrons \citep[e.g.,][]{markoff05}. Such models have been used with success to fit the data of some BH, and particularly GX 339$-$4 and XTE J1118+480 
\citep{markoff05,maitra09,Nowak11}. It should be noted, however, that they obviously rely on the presence of a compact jet and therefore can only be valid when/if a compact jet is present.

The capacities of the current high energy missions, albeit excellent, do not permit today an easy discrimination between the different families of models. This has nicely been illustrated in the specific case of Cyg X-1 through the 0.8--300\,keV spectral analysis of the Suzaku/RXTE observations \citep{Nowak11}. Different and complementary approaches and arguments can, however, be used to try and identify the most likely origin for the hard power-law tail. The fundamental plane between radio luminosity and X-ray luminosity \citep[e.g.,][]{Corbel03} has originally been used as an argument for a (small) contribution of the jet to the X-ray domain. Our approach here was to first separate the radio data sets into different spectral states, based on a model-independent classification.  It was only because of this separation that could we show that a hard polarized tail is present in the LHS, with a high polarization fraction, together with a radio jet. The very high level of polarization can only originate from optically thin synchrotron emission (see below), and implies a highly ordered magnetic field similar to that expected in jets. The synchrotron tail coincides with thermal Comptonization at lower energies and with a radio jet. The presence of all these features is qualitatively compatible with the multi-component emission processes predicted by compact jet models such as those of \citet{markoff05}.  While polarization of the (compact) jet emission has already been widely reported in the radio domain \citep[][and references therein]{brocksopp13}, a thorough study of the multi-wavelength polarization properties of Cyg X-1 has only recently been undertaken by \citet{russell13}. These authors, in particular, showed that the multi-wavelength spectrum and polarization properties of Cyg X-1 are quantitatively compatible with the presence of a compact jet that dominates the radio to IR domain and is also responsible for the MeV tail. A similar conclusion was drawn by \citet{malyshev2013} under the assumption that ``the polarization measurements were robust'' and that therefore the MeV tail could only originate from synchrotron emission. This pre-requisite is reinforced by our refined and spectral-state dependent study of the polarized properties of Cyg X-1 at 0.3--1\,MeV.  A polarized signal is indeed suggested by the 400--2000\,keV LHS data and the large degree of polarization measured in \S\ref{sec:res-polar} implies a very ordered magnetic field. Assuming the magnetic field lines are anchored in the disk implies that the $\gamma$-ray polarized emission comes from close to the inner ridge of the disk where the magnetic energy density is  highest. This constraint is difficult to reconcile with a spherically shaped corona medium, where the magnetic field lines are more likely to be tangled and where the polarized component necessarily comes from the outer shells of the corona where the optical depth is the lowest.  We, therefore, conclude that \textbf{in the LHS} the 0.4--2 MeV tail detected with \integral/IBIS is very likely due to optically thin synchrotron emission, and that this emission comes from the detected compact jet.

\subsection{Compatibility of the Cyg X-1 hard state parameters with synchrotron emission}
The synchrotron spectrum of a population of electrons following a power-law distribution, $dN(E)\propto E^{-p} d(E)$, where $p$ is the particle distribution index, can be approximated by a power-law function over a limited range of energy, i.e., $F_{E_1,E_2}(E)\propto E^{-\alpha}$ over $[E_1,E_2]$ \citep{rybickilightman}.  In the case of compact jets, two main domains are usually considered: the optically thick regime with $\alpha\lesssim0$ and the optically thin regime with $\alpha>0$. The break between the two regimes is typically in the infrared band \citep[e.g.,][]{Corbel02a,rahoui11,russell13,corbel13}.  The jet synchrotron spectrum is optically thin towards higher energies, and here $\alpha=(p-1)/2$.  The jet emission becomes negligible in the range from the optical to hard X-rays when compared to the contribution from other components such as the companion star, the accretion disk, or the corona. Our results indicate that it again dominates above a few hundred\,keV \citep[see also][for a multi-components modeling of the multi-wavelength Cyg X-1 spectrum]{russell13}.  

The degree of polarized emission expected in the optically thin regime is $\Pi=\frac{p+1}{p+7/3}$ \citep{rybickilightman}. Since the flux spectral index $\alpha=\Gamma-1$, where $\Gamma$ is the photon spectral index, it is possible to deduce $p$ from the measured spectral shape.  With $\Gamma=1.4_{-0.3}^{+0.2}$ (\S\ref{sec:hardstate}) we obtain $p=1.8_{-0.7}^{+0.4}$ and thus expect $\Pi_{\mathrm{expected}}=67.8_{-5.5}^{+2.7} \% $.  This value is consistent with the value measured in the LHS ($\Pi= 75 \pm 32\% $).

We determine a polarization angle of $\sim$$40^\circ$ compatible with the SPI results (JRC12). This angle, however, differs by about $60^\circ$ from the polarization angles measured in the radio and optical \citep[][and references therein]{russell13}. Assuming that the $\gamma$-ray polarized component comes from the region of the jet that is closest to the launch site (i.e., the inner ridge of the disk and/or the BH), the change in polarization angle implies that the field lines in the $\gamma$-ray emitting region have a different orientation than further out in the jet. This could be indicative of a relatively large opening angle at the basis of the jet \citep[as sketched in Fig.~3 of][]{russell13}. Alternatively the large offset could simply be due to strongly twisted magnetic field lines close to the accretion disk.

\section{Summary} 

We have presented a broad band 10\,keV--2\,MeV spectral analysis of the microquasar Cyg X-1 based on about 10 years of data collected with the JEM-X and IBIS 
telescopes onboard the \integral\ observatory. We have used the classification criteria of \citet{grinberg2013} to separate the data into hard (LHS), intermediate (IS), 
and soft (HSS) states.  We have studied the  radio behavior of the source associated with these states as observed with the Ryle/AMI radio telescope at 15~GHz.

\begin{itemize} 
\item The $\leq$400\,keV emission is well represented by Comptonization spectra with further reflection off the accretion disk.  In the LHS and IS, the Comptonization 
process is thermal and optically thick, i.e., the corona has an optical depth $\tau>1$ and the electrons have a Maxwellian distribution. As a result, a clear high-energy 
cut-off is seen in the spectra. In the HSS the situation is not that clear, although a cut-off is also necessary to give a good representation of the broad band spectrum.  

\item A clear hard tail is detected in the LHS when also considering the 0.4--2\,MeV data. This high energy component is well represented by a hard power-law with no 
obvious cut-off. The detection of the hard tail is compatible with earlier claims of the presence of such a component in spectra of Cyg X-1. We show that this component 
is variable within same state, as is seen when considering \integral\ observations from two arbitrary epochs but the same state.

\item In the radio domain, the 15\,GHz data show a definite detection with averaged flux densities of respectively $\sim$13 and $\sim$15\,mJy in the LHS and IS, 
compatible with the presence of a compact jet in those states. No persistent radio emission is detected in the HSS, implying the absence of a compact radio core.

\item In the LHS, we measure a polarized signal above 400\,keV with a large polarization fraction ($75\pm32\%$). This high degree of polarization and the polarization 
angle ($40\fdg0\pm14\fdg3$) are both compatible with previous studies by us and others. We obtain non-constraining upper limits on the polarization fraction in the 
IS and HSS, which have significantly lower exposure.

\item The high degree of polarization of the hard tail can only originate from synchrotron emission in an highly ordered magnetic field.  The demonstrated 
presence of radio emission in the LHS points towards the compact jet as the origin for the 0.4--2\,MeV emission, corroborating earlier theoretical and 
multi-wavelengths studies \citep{malyshev2013,russell13}.  Our spectral state-resolved and multi-wavelength approach therefore further confirms the 
conclusion presented in the earlier, non state-resolved studies based on \integral\ data only \citep{laurent11, jourdain12}.

\item We increased the total \integral\ exposure time and, in particular, nearly doubled the amount of data taken in the HSS.  We, however, still do not 
reach strong constraints on the polarization fraction in this state, even if we showed that the hard tail is much fainter.  Provided the source does not 
change state, we have gotten \integral\ time approved to further increase the exposure time in the HSS, which will allow us to obtain tighter constraints 
on the polarization signal in this very state.
\end{itemize}

\begin{acknowledgements}
This paper is based on observations with INTEGRAL, an ESA project with instruments and science data centre funded by ESA member states (especially the PI countries: Denmark, France, Germany, Italy, Switzerland, Spain) and with the participation of Russia and the USA.

We acknowledge S. Corbel, R. Belmont, J. Chenevez, and J.A.\ Tomsick for very fruitful discussions about several aspects presented in this paper. J.R. acknowledges funding support from the French Research National Agency: CHAOS project ANR-12-BS05-0009 (\texttt{http://www.chaos-project.fr}), and from the UnivEarthS Labex program of Sorbonne Paris Cit\'e (ANR-10-LABX-0023 and ANR-11-IDEX-0005-02). This work has been partially funded by the Bundesministerium f\"ur Wirtschaft und Technologie under Deutsches Zentrum f\"ur Luft- und Raumfahrt Grants 50\,OR\,1007 and 50\,OR\,1411. V.G.\ acknowledges support provided by NASA through the Smithsonian Astrophysical Observatory (SAO) contract SV3-73016 to MIT for Support of the Chandra X-Ray Center (CXC) and Science Instruments; CXC is operated by SAO for and on behalf of NASA under contract NAS8-03060. We acknowledge the support by the DFG Cluster of Excellence ``Origin and Structure of the Universe''. We are grateful for the support of M. Cadolle Bel through the Computational Center for Particle and Astrophysics (C2PAP).
\end{acknowledgements}


\newpage

\newpage

\newpage

\newpage

\newpage 

\end{document}